%% file: wp-main.tex
\documentclass[10pt]{bmcart-tuned}

\RequirePackage{etoolbox,amsmath,amssymb,float,hyperref,cleveref}

\startlocaldefs

\input{./wp-meta-data.tex}

\input{./wp-template.head.tex}

\input{./wp-template.config.tex}

\input{wp-article-header.tex}

\endlocaldefs

\begin{document}

\input{wp-hard-cover.tex}

\input{wp-article-first-page.tex}

\section{Introduction}

\begin{submittedText}\submittedTextNote{JAIS March 2016}
\lettrine[]{M}{any}
remark that ``networks are everywhere!'' \citep{latour2011_on_ANT,newman_ety_al2006_a_book_on_network_dynamics,dorogovtsev2002evolution,strogatz2001_on_complex_networks,cohen2002all_w_a_net}. Examples of networks recurrently modeled by scholars are the Internet and other infrastructures, social, political and economic networks. Also  neural, inter-organizational, scientometric, and  text-representational networks among many others. As the network paradigm becomes scientifically relevant across disciplinary boundaries, scholars recurrently turn to network science -- an emerging field, that like statistics,  permeates a wide range of traditional disciplines \citep{brandes_et_al2013what_is_network_sciences}. 

\end{submittedText}

\begin{submittedText}\submittedTextNote{JAIS March 2016} 
 Network science, as the study of the collection, management, analysis, interpretation, and presentation of relational data, provides scholars across disciplines with both theory and methods to deal with the increasing availability of relational datasets \citep{brandes_et_al2013what_is_network_sciences}.  Given the trans-disciplinary nature of our field \citep[see][]{galliers2003change} and as information systems become increasingly networked and interconnected \citep{henfridsson_and_bygstad2013generative,ciborra_et_al2000_dynamics_digital_infrastructures}  the network paradigm is gaining relevance in the discipline. 
\end{submittedText}

\ifbool{shortVer}{}{
\begin{submittedText}\submittedTextNote{JAIS March 2016}

The emergence of social networking websites (e.g., Facebook, LinkedIn, and Twitter among many others) and web-based open-source software repository/hosting services (e.g., GitHub, Bitbucket and Black Duck Open Hub),  materialized big datasets capturing interactions both between millions of users and millions of systems --- such broad data sets became ``a gold mine''  for  social scientists to leverage network-based theory and methods \citep[see][]{borgatti_et_al2009sna_in_social_sciences}.
The impact of network science to the IS discipline had been already recognized \citep[e.g.,][]{sundararajan2013research,johnson2014emergence,gu2014research,haturvedi_et_al2011}.  So far, the phrase social network analysis (``SNA'') was the most used label. To our view, this suggests that our discipline overlaps more with network science by its methodological foundations than by theory. In other words, our discipline seems more interested in network analysis than in network theory. 
\end{submittedText}
}

\begin{submittedText}\submittedTextNote{JAIS March 2016}
In this research, we explore one of the most important theoretical concepts in network science -- homophily.  The term homophily (etymologically from Greek; \textit{homoios}: equal, similar; \textit{philia}: friendship, love, affection)  describes the relational tendency of individuals to associate and bond with similar others. We test such “love of the same” principle in peer-review networks by examining the evolution of the Linux kernel\footnote{To best of our knowledge, Linux is the most studied software development project.}.
\end{submittedText}

\begin{submittedText}\submittedTextNote{JAIS March 2016}
The principle of homophily suggests that actors tend to establish ties with similar others. Homophily has been previously explored in information systems research within multiple settings 
\ifbool{shortVer}{
\citep{gallivan_and_ahuja2015co}
}{
\citep[e.g.,][]{gallivan_and_ahuja2015co,gu2014research,putzke2010evolution,hovorka_and_larsen_2006enabling}	
}.  Among other examples, while investigating the adoption of a large-scale IT system across multiple sites in New York State, \citet{hovorka_and_larsen_2006enabling} confirmed that organizational similarity influenced the willingness of organizations to establish
and maintain communication ties.   In a scientometric study examining the evolution of co-authorships in top IS journals, 
 \citet{gallivan_and_ahuja2015co} found significant effects of homophily related to gender, proximity, and geography. IS
scholars worldwide exhibit a stronger preference for collaborating with co-authors of the same sex and those who
attended the same Ph.D. program.  More recently, \citet{chipidza2016our} found homophily related to gender, geography, and graduation year in the co-authorship network of the IS Senior Scholar Basket of 8. Among other 	circumstances, homophily was addressed by IS scholars within the context of multi-player on-line games  \citep{putzke2010evolution},  virtual investment-related communities \citep{gu2014research}, online shopping \citep{gaskin2010bypassing}, and  open-source communities \citet{hu2009discovering,Hu2012526}.

\end{submittedText}

\begin{submittedText}\submittedTextNote{JAIS March 2016}

This paper builds on the interdisciplinary tradition of network science -- We overlap network theory, network analysis, open-source and information systems research. At the theoretical level, the principle of homophily is our principal theoretical interest. Even if this research is embedded in a larger project that aims at making sense of \emph{who reviews who in the Linux kernel?}, the more workable initial research question provided the framing: ''Does the code reviews in the Linux Kernel tend to be homophilous``?. Or in other words ``Does the contributions to the Linux kernel tend to be reviewed by people who are similar to the original contributors?''.
Assessing homophily is important as  innovative organizations should ideally disclose low levels of homophily  as  successful  collaboration  requires  complementary over substituting actors \citep[][pp. 125]{desouza2011intrapreneurship}. In addition, low levels of homophily facilitate communication among very distinct actors \citep{rogers1976,granovetter1973strength_of_weak_ties} and ease the  recombination of ideas  across diverse
areas of the knowledge possessed in the team \citep{fleming_et_al2007on_brokering_vs_cohesive_creativity,west2002on_creativity_and_innovation_in_work_groups}. 
At a more empirical level, the paper joins to the group of existing multidisciplinary case studies that have examined the Linux kernel and its development 
\ifbool{shortVer}{
\citep[e.g.,][]{Shaikh2017, homscheid2016firm, Ermann11,hertel_et_al2003on_linux_developers_motivations}
}
{
\citep{Ermann11, homscheid2016firm, Israeli10, lee_and_cole2003oss_model_of_knowledge_creation, Palix14, WangYu13,lee_and_cole2003oss_model_of_knowledge_creation,moon_and_sproull2002essence,hertel_et_al2003on_linux_developers_motivations,Shaikh2017}
}.
The empirical longitudinal analysis concentrates on five kernel subsystems (i.e., arch, drivers, fs, kernel and net)  and their evolution from 2006 to 2014. In total 45 directed and weighted peer review networks are analyzed. The distributed version control system used in the Linux kernel development provides the source of empirical data.  The empirical analysis relies on descriptive statistics and different network indices, preserving the directed and weighed nature of peer review networks.

\end{submittedText}

\begin{submittedText}\submittedTextNote{JAIS March 2016}
\end{submittedText}

\section{Theoretical background} \label{section:theory}

\subsection{Network science and homophily} \label{subsection:ns_and_homophily}

\begin{submittedText}\submittedTextNote{JAIS March 2016}
Network science provides theory and methods that are particularly suited to analyze and explain phenomenas where dependence is observed both between and within variables \citep{brandes_et_al2013what_is_network_sciences}. Very often, IS scholars make strong assumptions about the independence of observations to access to standard theory in inferential statistics \citep{mingers2004_pushing_for_critical_realism,lindberg_et_al2013computational}. In network science, dependencies between and within variables, are not undesirable nuisances or defects to be leaved out by the study design, but they often constitute the actual research interest.  As in spatial statistics, observations are not assumed to be independent of each other but are explicitly set up to have structure. Such dependence structure between observations is often what network scientists are after \citep{brandes_et_al2013what_is_network_sciences}. By taking such stance, the behavior of different information systems users can be influenced by ties among them (e.g., friendship, work-relationship); or the successful implementation of an IT system can be influenced by many relational factors (e.g., social networks, structure of value nets, systems interoperability).
\end{submittedText}

\begin{submittedText}\submittedTextNote{JAIS March 2016}
One of the most fundamental characteristic of network theory in social sciences is the focus on relationships among actors as an explanation of actor and group outcomes \citep{borgatti_and_halgin2011_on_network_theory,borgatti_et_al2014on_sna_criticism}. The principle of homophily (i.e., that actors tend to establish ties with similar others in a group) is, therefore, central to network science. Such positive relationship between the similarly of two actors in a group  and the probability of a tie between them was one of the first features early noted by network scientists \citep{freeman1996sna_antecedents}. As pointed out in the seminal homophily-review by \citet{mcpherson2001_a_review_of_homophily}, structural sociologists have studied homophily in relationships that range from the closest ties of marriage \citep{kalmijn1998intermarriage} and the strong relationships of ''discussing important matters''\citep{marsden1987_homophily_in_discussion} and friendship \citep{verbrugge1977homophily_in_friendship} to the more circumscribed relationships of career support at work \citep{ibarra1992homophily_in_the_firm} to mere contact \citep{wellman1996personal}, ``knowing about'' someone
\citep{hamptonand_wellman2001_homophily_in_aquiantance} or appearing with them in a public place \citep{mayhew1995_homophily_in_appearing_together}. 
\end{submittedText}

\begin{submittedText}\submittedTextNote{JAIS March 2016}
 Research on the patterns of homophily is remarkably robust over various types of relations \citep{mcpherson2001_a_review_of_homophily,kossinets_and_watts2009origins_of_homophily}. Studies that measured multiple forms of relationship
have shown that the patterns of homophily tend to get stronger as more types of relationships  exist between two actors -- multiplex ties generate greater homophily than simplex ties \citep{fischer1982dwell,fischer1982we,hristova2014keep_homophily_in_multiplex_online_and_offline,renoust2014entanglement_in_multiplex_two_data_sources,grossetti2007_reproduces_fischer_work_in_france,haythornthwaite1998_on_multiplex_homophily}. Evidence of the homophily tendency crossed domains, in scientific fields (e.g., physics and biochemical networks) the same tendency is known as assortative mixing \citep{Croft08, Newman03b, Peng15}. Particularly in social sciences, evidence was found that ``similarity breed connection'' with regard to many %
characteristics such as gender, ethnicity, age, religion, education, occupation, social class hierarchy, geography, family, organizational affiliation, network positions, attitudes, abilities, believes, and  aspirations among others  (see \citealp{mcpherson2001_a_review_of_homophily,brass_et_al2004multi_level_perspective}, chapter 6 in \citealp{Croft08} and chapter 4 in \citealp{easley2010networks} for exhaustive reviews).
\end{submittedText}

\begin{submittedText}\submittedTextNote{JAIS March 2016}
Both classical and computational social science studies\footnote{See 
\citet{lazer_et_al2009on_computational_social} for a discussion on the emergence of data-driven ``computational social science''  in general and \citet{lindberg_et_al2013computational} on its application to open-source in particular} 
 recurrently confirmed that homophily bounds social networks \citep{bakshy_et_al2015homophily_in_facebook,colleoni_et_al2014homophily_in_twiter,aral_et_al2009homophily_and_platform_adoption}. However, there are non-confirmatory studies as well. Three very recent studies on the open-source software domain did not confirm the expected pattern of homophily. In a recent exploratory case study, \citet{teixeira_et_al2015lessons} investigated homophily in the joint-development of the OpenStack -- an open-source infrastructure for big data co-developed by hundreds of organizations and thousands of developers (e.g., Rackspace, Canonical, IBM, HP, Vmware, Citrix, Intel and AMD  among others). Contrary to expected, the analysis of a complex network on ``who codes with who'' revealed that developers did not tend to work with developers from the same firm in the ecosystem. There was a relational tendency among OpenStack developers  to work with developers from other companies (often competitors). A similar study by \citet{linaaker_et_al2016_coopetition_in_Hadoop} investigated the Hadoop -- open-source distributed storage and processing technologies for big data joint-developed by extensive network of participants (e.g, Cloudera, Yahoo!, Facebook, Twitter, LinkedIn, Jive, Microsoft, Intel and Hortonworks among others). The results pointed out that developers affiliated with competing firms collaborate as openly as the ones affiliated with non-rivaling firms do.   In such dyad of recent  studies, the theoretically expected homophily regarding company affiliation was not observed in open-source communities as in other social networks.
 \ifbool{shortVer}{}{
 \footnote{As a side note, another related study exploring gender bias in open-source software development was recently pre-archived by \citet{terrell_et_al2016gender}. By mining by GitHub and Google+, they found out that when a woman's gender was not obvious from her name or photo in the GitHub profile, their patch would be slightly accepted more often than a patch from a man. However, as the study was not yet published in a peer-reviewed outlet, we opted to not built upon it.}
 }
 Such a difference between the patterns of homophily in open-source communities and the patterns on homophily in other social networks motivates further examinations. After all,  understanding of the dynamics of homophily, can lead to more effective reward structures and more creative collaboration structures \citep{gallivan_and_ahuja2015co}.
 \end{submittedText}

 \improvement{Shall we mention here the usual criticism of network science. See \url{http://dx.doi.org/10.1108/S0733-558X(2014)0000040001}}

\subsection{Peer review networks and open-source software} \label{subsection:pr_n}

\begin{submittedText}\submittedTextNote{JAIS March 2016}
Peer review is an essential element of modern science. Although the history of scientific peer review traces back to the classical antiquity, the roots of the contemporary institutional arrangements are located in the 18th century process that was first adopted by \textit{Philosophical Transactions}, the first journal exclusively devoted only to science~\citep{Bornmann11, Spier02}. While definitions, interpretations, and opinions tend to vary, quality control is usually still seen to be the main purpose of scientific peer review.
\end{submittedText}

\begin{submittedText}\submittedTextNote{JAIS March 2016}
This fundamental trait is present also in different software development peer review practices.
\ifbool{shortVer}{}
{
\footnote{~The general terms peer review and  code review are used interchangeably in this paper. It should be emphasized that the
particular term software code inspection refers to a different, more encompassing software 
practice \citep{Fagan76, Rigby14}. These terminological issues do not affect the case study results, however.} 
}
This trait becomes evident by considering the theoretical attempts to define the overall goal of reviewing or inspecting artifacts written by others. According to some authors, the explicit objective is to find errors \citep{Fagan76}. While other definitions enlarge the theoretical scope, for instance by including the goal to find deviations from specifications \citep{Aurun02}. 
Recent research explicitly addressing the benefits of code reviews pointed out that developers spend 10-15 percent of their time to find defects, share knowledge, build a sense of community, and maintaining quality \citep{Bosu_et_al2017}.

\end{submittedText}

 \begin{submittedText}\submittedTextNote{JAIS March 2016}
Many relational concepts from network theory have been observed within open-source software (OSS) collaboration and peer reviewing networks.
On one hand,  empirical network tendencies such as network centralization, and the associated theoretical concepts such as the core and periphery, have long been adopted to successfully describe relational socio-technical OSS collaboration \citep{Bird11, Crowston06, Toral10}. On the other hand, actors' attributes such as the activity, experience, and seniority, have been observed to have an impact upon OSS peer review networks and their efficiency \citep{Baysal13, Bosu14b, Rigby14}. 
\end{submittedText}

\ifbool{shortVer}{}{
\begin{submittedText}\submittedTextNote{JAIS March 2016}
Recent research has provided a convincing amount of empirical evidence particularly in terms of different OSS peer review metrics \citep{Baysal13, Bettenburg15, Rigby14}. Analogously to the more widely used scientific counterparts, there are, of course, a~number of limitations related to such quantitative indicators \citep{Johnson98}. Nevertheless, the empirical research has been able to show substantial differences in the practices, and, hence, the inefficiencies between different open-source projects. In terms of the Linux kernel, in particular, rejections seem to arrive quickly, but decisions on acceptance can have considerable time lags \citep{Bettenburg15}, to only point out one telling empirical result that provide a sufficient rationale to consider the peer review process in the Linux kernel from a network perspective. 
\end{submittedText}
}

\begin{submittedText}\submittedTextNote{JAIS March 2016}
This paper contributes by 
exploring %
homophily as an important theoretical tendency in open-source peer review networks. As posited by \citet{Robins07}, understanding the theoretical reasons for a hypothetical network topology is important to ensure further research advances, and to allow the potential for further theorizing. %
To best of our knowledge, this is the first paper exploring homophily by longitudinally investigating peer-review along with the development of a high-networked information system (i.e., Linux). 
\end{submittedText}

\section{Research design}\label{section: research design}

\begin{submittedText}\submittedTextNote{JAIS March 2016}
This research was conducted as an empirical case study design informed both by methodological notes on case study research 
\citep[][]{eisenhardt1989building}
and social network analysis  \citep{wasserman_and_fast1994,howison2011validity,contractor2006testing}. We assess the peer review practices of a large open-source software project driven by our interests in network theory (i.e. homophily).  Our research design also reflects extant knowledge in peer review practices and processes of open-source communities in general and Linux in particular. 

\end{submittedText}

\subsection{Case selection} 

From the empirical side, we motivate the case of Linux by pointing out that according to the empirical data, since the mid-2000s, the Linux kernel was developed by a network of with over eleven thousand distinct individuals. Many refer to Linux as the most successful, and the most important open-source project of all time. Also, in relation to existing literature, Linux is the most studied open-source project \citep{crowston2012free}. As much research in  Linux already exists, analyzing the peer review networks of Linux have a higher potential of integration with prior research. As pointed out in a recent critical review of research in open-source software,  ``investigating the phenomenon within a small and confined domain and gradually extending the validity of the results through replication is a much sounder approach rather than over-generalizing the results of a study to a broad domain without any theoretical justification or empirical evidence'' \citep{carillo_and_bernard2015many}. 

From the theoretical side, the Linux kernel is a particularly relevant case to observe the important question of firm engagement in OSS development. First of all, the commercial aspects provide also one motivation for homophily theory --  if the engaged affiliates from a company would be systematically reviewing each other, as has been suspected in some cases \citep{Baysal13}, the normative ideals of OSS peer reviews would be arguably violated.  
Second, as Linux is a mature project with \textit{de facto} hierarchies\citep[cf.][]{Crowston06,o_mahony_and_ferraro2007emergence_governance_oss,Shaikh2017} and such  hierarchies
apply to  the maintenance and patch submission  practices \citep{Kleen, Love05}  and policies  \citep{Kernel15a},  hierarchical roles are likely to impact homophilous  behaviour \citep[][]{dodds2003information,shen_and_monge2011connects}. In \citet{benkler2006wealth} terms, OSS communities often display a ``meritocratic hierarchy'', which does not hinge on employment authority as in traditional firms. When common social attributes are largely absent or obscured (e.g., age, race, educational background)
individuals could resort on project leadership, merits and reputation when deciding to "network "  with other developers. 
Third, the Linux kernel provides an important case to reflect upon the software development peer review practices against the scientific counterparts \citep{lee_and_cole2003oss_model_of_knowledge_creation}. The question is important already because scientific peer reviewing has long provided empirical cornerstones for many of the high-level network theories \citep{Merton68, Newman01, Peng15}. These three reasons provide an ideal frame to follow the interdisciplinary network science domain. While the research observes socio-technical networks, the perspective is not attached to any specific discipline in the larger social or technical research domains. 

\subsection{Conjectures} \label{subsec:conjectures}

\begin{submittedText}\submittedTextNote{JAIS March 2016}
The research question -- does the code reviews in the Linux Kernel tend to be homophilous? -- can be disaggregated into a few assertions. As the pursued methodology is based on descriptive statistics, these assertions should be understood as analytical research vehicles rather than strictly testable hypotheses. To make this difference explicit, the assertions are labeled as general conjectures of the underlying network theory. Besides attempting to gain some theoretical precision by reflecting prior expectations, these conjectures are also meant to control the common \textit{post hoc} rationalization that has often been argued to be a typical element in case studies \citep{Campbell75,bitektine2007prospective}. 
\end{submittedText}

When facing the empirical material, and to exploit homophily related to exogenous network elements (i.e., attributes of software developers modeled as network nodes), the following  conjectures can be stated, given two exogenous node attributes of interests: 

\begin{enumerate}[label=C$_{\arabic{enumi}}$, resume]
\item{\textit{Maintainers tend to review other maintainers.}}\label{h: maintainers and maintainers}
\item{\textit{Developers affiliated with a organization (i.e., company, university) tend to review other developers affiliated with the same organization.}}\label{h: affiliations}
\end{enumerate}

Maintainers are granted with a formal role within the Linux ``meritocratic hierarchy'' and they are expected to review contributions to the various parts of the Linux kernel that they maintain. Our second exogenous attribute of interest, affiliation, results from an employment or contractual relationship with an organization that commits resources to the development of the Linux kernel. While being a maintainer reflects meritocracy as a value of open-source communities, affiliation reflects the commitment of an individual to an organization. Such operationalization  aligns with prior research relating homophily with hierarchy \citep[see][]{shen_and_monge2011connects,svskerlavaj2010patterns} and homophily with organizational affiliation \citep[see][]{teixeira_et_al2015lessons,kim2007alliances}.

Finally, the research design allows to generalize these conjectures with the following two corollaries.

\begin{enumerate}[label=C$_{\arabic{enumi}}$, resume]
\item{\textit{The results are similar across the main subsystems}.}\label{h: subsystem similarity}
\item{\textit{The results are similar across the annual subsamples}.}\label{h: annual similarity}
\end{enumerate}

\subsection{Data collection and analysis}\label{section: data}

\begin{submittedText}\submittedTextNote{JAIS March 2016}
To extract the desired peer review data, we ``borrowed'' extensive knowledge from the Mining Software Repositories (MSR) field that provides many methods and tools to analyze the rich trace data available in software repositories. After gaining insights on Linux and its development processes, we defined keywords and used pattern matching techniques with regular expressions\footnote{Due to lack of space, we opted to not detail the complex and fine-grained details of our pattern matching procedures that mined the Linux software repository (orchestrated with Git).}
to obtain the desired relational data provided by the software repository orchestrating the development of Linux. 
\end{submittedText}

\ifbool{shortVer}{
Guided by cross-disciplinary methodological notes that overlap the knowledge on the Mining of Software Repositories with the analysis of networked digital trace data \citep[e.g.,][]{howison2011validity,Bird07,Bettenburg15}, we modeled  the patch's delivery path \citep[see][]{Kernel15a} distinguidhing between authors and reviewers \citep[see][]{Bettenburg15}. From  naturally occurning traces of the Linux development history (signatures that credit the contributors of the Linux kernel) we could build networks of who reviewed who by analysing each code-contribution from the time it is submitted until the time it 'lands' and merges into the official Linux code base. 
}{

\begin{submittedText}\submittedTextNote{JAIS March 2016}
Elucidating our keywords-based data mining process, the following cut excerpt shows a typical commit message structure used in the Linux kernel development:

\begin{scriptsize}
\begin{verbatim}
    commit 77ea8c948953a90401e436e7c05973b2d5529804
    Author: Dave Young <dyoung@redhat.com>
    Date:   Fri Dec 20 18:02:22 2013 +0800

    [...]
    
    Signed-off-by: Dave Young <dyoung@redhat.com>
    Acked-by: Borislav Petkov <bp@suse.de>
    Tested-by: Toshi Kani <toshi.kani@hp.com>
    Signed-off-by: Matt Fleming <matt.fleming@intel.com>
\end{verbatim}
\end{scriptsize}

\end{submittedText}

\begin{submittedText}\submittedTextNote{JAIS March 2016}
The first line in this standard but customizable format includes a computational hash that uniquely identifies all commits. Author and date are shown in the next two lines, after which the actual commit message body follows. This format allows to build a relational empirical dataset based on the simple analytical distinction between authors and reviewers \citep{Bettenburg15}, that is, between those individuals who authored a commit and those who are marked with three visible keywords in the commit message body. All other potential roles are excluded in this paper.\footnote{~A few additional clarifications are required regarding the \texttt{git} version control system used for the Linux kernel development. First, the vanilla source code tree is used \citep{Kernel15b}. Second, all considered authors are also \texttt{git}-authors (and not committers), meaning that those individuals who wrote a patch are observed rather than those who committed the patch on behalf of the authors. Third, the time period endpoints were, however, specified according to the so-called commit date that records a timestamp at which a commit was made to include a patch in the main tree. Fourth, merges between branches are excluded. Last, data for the subsystems refer to the corresponding directories in the root of the source code collection.} 
\end{submittedText}

\begin{submittedText}\submittedTextNote{JAIS March 2016}
The most important keyword is related to the so-called signature line that is visible also in the above excerpt. It signifies all participants who were involved in the development of a patch, and all those who were in the patch's delivery path \citep{Kernel15a}. %
As the signatures leave visible accounting traces to the commit history, also the meritocratic OSS development credits can be thought to follow these traces, for instance. There are, moreover, a couple of additional keywords that are closely related to peer review but without the supposed juridical aspects; namely, acknowledged (who approved and reviewed informally) and reviewed (who reviewed with sufficient rigor) \citep{Kernel15a}. While there is a specific keyword for code reviews, also the other two keywords can be interpreted to exhibit elements of peer review. 
\end{submittedText}. Du
\improvement{This sections is not so clear !!}

\subsubsection{Statistical procedure}

\begin{submittedText}\submittedTextNote{JAIS March 2016}
The empirical sample is based on rather simple regular expression routines wrapped around the described three keywords. These expressions are enumerated in Table \ref{tab: review keywords}. %
\end{submittedText}

\begin{submittedTable}[JAIS March 2016]
\begin{table}[H]
\centering
\caption{Regular expressions for peer review keywords}
\label{tab: review keywords}
\begin{threeparttable}
\begin{small}
\begin{tabular}{p{4cm}p{4cm}p{4cm}}
\toprule
Signed & Acknowledged & Reviewed \\
\midrule
\texttt{signed}$\bullet$\texttt{by}$\odot$ &
\texttt{acked:} &
\texttt{reviewed:} \\
\texttt{signed}$\bullet$\texttt{of}$\bullet$\texttt{by}$\odot$ &
\texttt{acked}$\bullet$\texttt{off}$\bullet$\texttt{by}$\odot$ &
\texttt{reviewed}$\bullet$\texttt{by}$\odot$ \\
\texttt{signed}$\bullet$\texttt{off}$\bullet$\texttt{by}$\odot$ &
\texttt{acked}$\bullet$\texttt{by}$\odot$ \\
\bottomrule
\end{tabular}
\end{small}
\begin{scriptsize}
\begin{tablenotes}
\item[]{The table enumerates the used regular expressions. These were matched from the beginning of each commit message (body) line by using lower case Unicode letters; $\bullet$ denotes either a space or a hyphen, while $\odot$ refers to a space, colon, or semicolon. Each expression is furthermore directly followed by \mbox{$\bigstar$\textless$\bigstar$@$\bigstar\centerdot\bigstar\circledast$}, where $\bigstar$ represents a wild card that matches at least one undefined character until the subsequent character, the symbol $\centerdot$ captures either a dot or a comma, and $\circledast$ is the greater-than sign or empty.}
\end{tablenotes}
\end{scriptsize}
\end{threeparttable}
\end{table}
\end{submittedTable}

\begin{submittedText}\submittedTextNote{JAIS March 2016}
The actual data collection procedure can be briefly described as follows. For each commit in each subsystem in each year, the regular expression routines were used to identify individuals in relation to a given commit author. Since e-mail addresses tend to vary \citep{Bird11}, both authors and reviewers were identified by their real names. Moreover, the common distance metric of \citet{Levenshtein66}, which seems reasonable for the purpose \citep{Zangerle13}, was used to account typing mistakes and other small inconsistencies in individuals' names.\footnote{~Specifically: if $s_1$ and $s_2$ denote two names with lengths $\vert s_1 \vert$ and $\vert s_2 \vert$, and $L(x, y)$ is \citeauthor{Levenshtein66}'s \citeyearpar{Levenshtein66} distance, a similarity score was computed from $\delta = 1.0 - L(s_1, s_2) / \max\lbrace \vert s_1 \vert, \vert s_2 \vert \rbrace$ \citep{Christen12}. If $\delta$ exceeded a manually constructed threshold value of 0.85, two names were taken to refer to the same individual. As the exceedances were further evaluated manually, some ambiguities must be acknowledged, but the benefits are still seen to exceed the costs. Moreover, it is worth remarking that a comparable approach has also been used~previously \citep{Rigby14}.} A few additional corrections were made to account some obvious human errors. If a keyword was accidentally followed by another similar keyword, for instance, the other one was removed. It should be emphasized that the routines are still more stringent than plain text searches in that a keyword line is only qualified when it contains at least partially valid entries for a real name and e-mail address. Almost all commits satisfied these minimum requirements.
\end{submittedText}

}
\change{Very detailed -- turns against us}

\begin{submittedText}\submittedTextNote{JAIS March 2016}

\ifbool{shortVer}{
Authors and reviewers were identified by their names as they appear in the repository commit logs. Moreover, the common distance metric of \citet{Levenshtein66}, which seems reasonable for the purpose \citep{Zangerle13}, was used to account typing mistakes and other small inconsistencies in individuals' names.
}{}
Peer review relationships were modeled as  weighted adjacency matrices, $\mta_1, \mta_2, \ldots$, the dimensions of which are based on the approximately unique individuals who have either authored or reviewed commits in a given subsystem during a given year. A matrix $\mta_i$ contains the conventional algebraic representation of a network in which individual elements represent the presence or absence of a link between two nodes according to whether an element $a_{xy} \in \mta_i$ is positive. No  truncation \citep{Hong11} or dichotomization \citep{Crowston06, Conaldi13} was applied to the weights. The resulting networks are directed already by the nature of peer review, and, hence, the matrices are asymmetric by definition. 
\end{submittedText}

\begin{submittedText}\submittedTextNote{JAIS March 2016}
In the context of this paper $\mta_i$ represents more specifically an abstract composite in which columns denote authors and rows individuals who have peer reviewed commits. For example, a value two for an element $a_{31}$ means that the third observed developer has peer reviewed two commits of an author whose commit records are given in the first column, irrespective whether the third column sum is zero, that is, regardless whether the third developer has been an author himself. It is worth noting that in the Linux kernel authors should always sign off their own commits. If the element $a_{11}$ would, thus, take a value three, for instance, the first observed developer would have authored three commits, and correctly signed them off with the compulsory signature. By implication, the number of signed commits per developer can be extracted directly from the diagonal of $\mta_i$. As is typical \citep{Robins13, Holland70}, however, all diagonal elements are excluded in the actual network analysis already due to the theoretical absurdity of a single person peer review.
\end{submittedText}

\begin{submittedText}\submittedTextNote{JAIS March 2016}
Regarding corporate affiliation, we adopted a strictly extrarelational approach: individuals are identified by their real names, while all affiliations are based on explicit data extraction from the domain names in individuals' e-mail addresses. 
\end{submittedText}

\begin{submittedText}\submittedTextNote{JAIS March 2016}
The actual identification was carried out semi-manually, to borrow a term used by \citet{Hamasaki13}. The frequent unique domain names were first examined manually in order to construct a suitable extraction scheme for each group. In general, the rightmost subdomain was considered as sufficient for the majority of cases (e.g., 't.jhun@samsung.com' identifies a developer's affiliation with Samsung). If a recurrently occurring subdomain was not instantly recognized, a visit was attempted with a web browser (e.g., search engines as well as the LinkedIn social network). If this visit did not provide new information, a search engine was used for detective purposes. If also this search failed, the subdomain was finally tagged as unaffiliated. 
\end{submittedText}

\begin{submittedText}\submittedTextNote{JAIS March 2016}
Finally, all maintainers are identified from a file supplied in the root of the source code collection \citep{Kernel15c}. Since the file has been updated numerous times, the identification was done with a twofold matching procedure: for all individuals in each subsystem in each year, the corresponding real names were searched from the first annual revision of the file, and from all subsequent patches to the file until the last commit in the given year. While this procedure ensures that annual changes in the maintainership are accounted for, no attempts are made to match the distinct code sections that maintainers are responsible for. In terms of %
\ref{h: maintainers and maintainers} this choice means that a maintainer whose responsibilities are located in the $i$:th subsystem is still classified as a maintainer even when he or she reviews commits to the $j$:th subsystem, for instance.
\end{submittedText}

\section{Results}\label{section: results}

 In order to analyze homophily in the peer-review networks of the Linux Kernel, we looked at the endogenous elements of the network (i.e., without attaching additional extra-relational attributes to the associated nodes or links).
 The corresponding prior expectations (i.e., the conjectures) are tested by means of fundamental network measures and descriptive statistics that capture the evolution of five kernel subsystems from 2006 to 2014. 
 
\subsection{Trend analysis}

\begin{figure}

     \subfloat[Number of nodes/developers\label{subfig:number-nodes}]{
         \centering
    \includegraphics[width=0.7\textwidth]{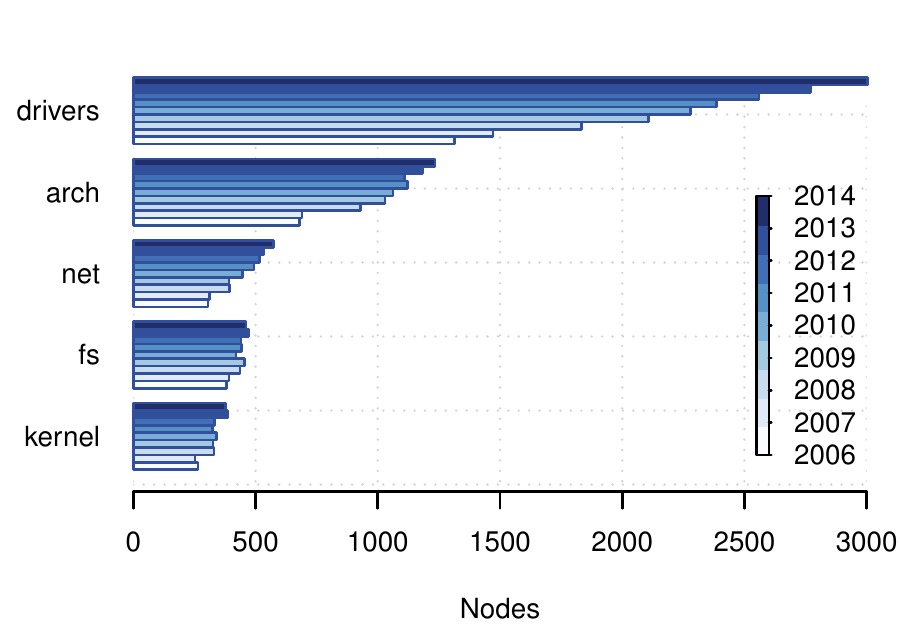}
    }
    \\ 
    \subfloat[Share of maintainers (\%) \label{subfig:maintainer-shares}]{        
         \centering
        \includegraphics[width=0.7  \textwidth]{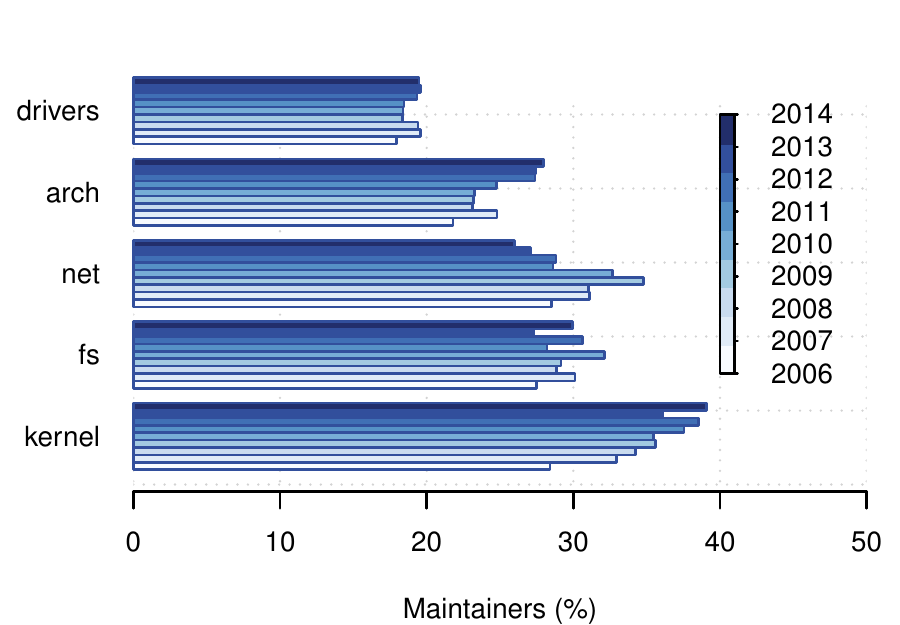}
     }
     \caption{Evolution trends}
     \label{fig:evolution-trends}
\end{figure}

The amount of individuals involved in peer review activities has indeed increased substantially in the Linux kernel during the past decade \citep{Bettenburg15}. As can be seen from 
Fig. \ref{subfig:number-nodes}
, the number of nodes has increased steadily within \texttt{drivers}, \texttt{arch}, and \texttt{net}. The \texttt{fs} and \texttt{kernel} subsystems seem to exhibit a slower rate of growth, however. As generally suspected by \citet{Toral10}, the reason may relate to the strong company presence in the former three subsystems, which are all closely related to hardware. On the other hand, \texttt{drivers} presumably garners also a large amount of volunteers and hobbyists working with consumer hardware.

\subsection{Maintenance as a foci of homophily}

\begin{figure}[hb]
 \centering
 \includegraphics[width=3.6in, height=4.2cm]{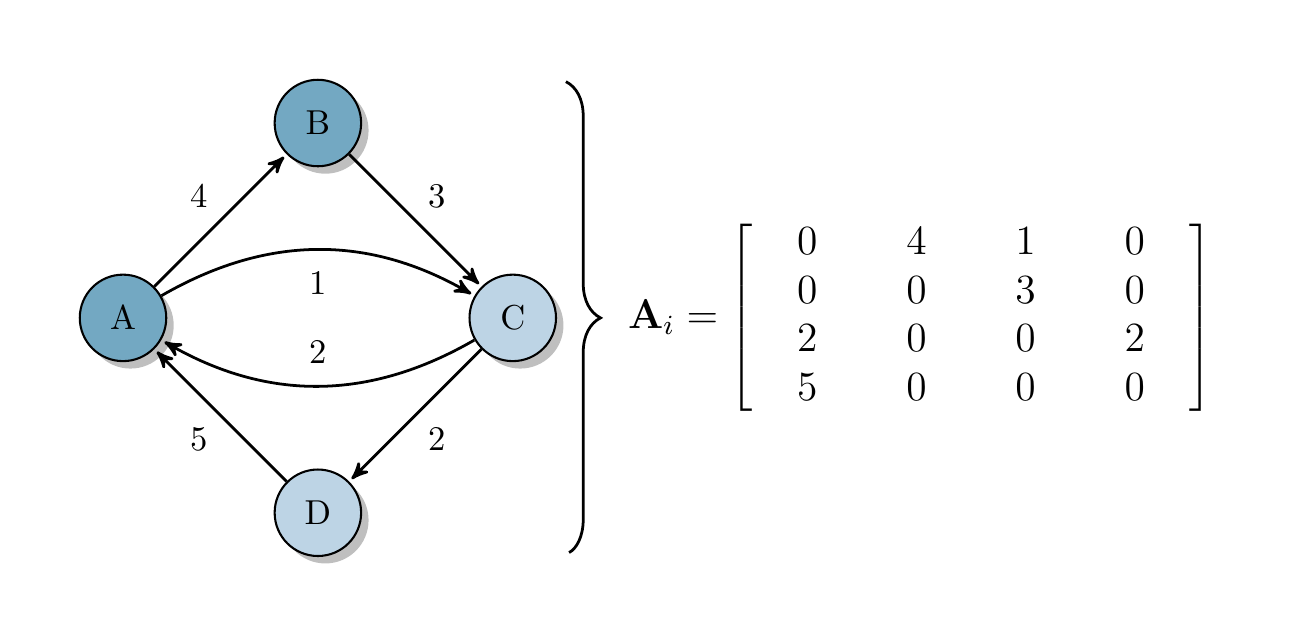}
 \caption{A small example network}
 \label{fig: example network}
\end{figure}

\begin{submittedText}\submittedTextNote{JAIS March 2016}
Guided by Blau's \citeyearpar{blau1977on_inequality_and_structure} established theoretical ideas, we employ a frequency-based analytic strategy \citep[][pp. 418]{mcpherson2001_a_review_of_homophily}  as previously employed by IS scholars to assess homophily in social networks \citep[e.g.,][]{gallivan_and_ahuja2015co,gu2014research}.  
In order to evaluate \ref{h: maintainers and maintainers}, we 
developed a simple custom index which presumes a tendency for maintainers to review other maintainers\footnote{~The group centrality index of \citet{EverettBorgatti05} was considered as an alternative, but the custom index was seen preferable as the group centrality measure is restricted to undirected and unweighted networks.}. For any maintainer, the index is defined as the ratio of reviews that have targeted maintainers to all reviews carried out by the node. (A value zero is reserved for those maintainers who have not reviewed at all.) Using Fig.~\ref{fig: example network} as an illustrative example network: if A and B are maintainers and the remaining two nodes denote normal developers, the corresponding review ratios for the two maintainers are $4 / 5 = 0.8$ and zero, respectively. The index is undefined for C and D since these two nodes lack the attribute flag for maintainership. The higher the value, the higher the degree of homophily.
\end{submittedText}

\begin{figure}[ht]
\centering
\includegraphics[width=0.95\textwidth]{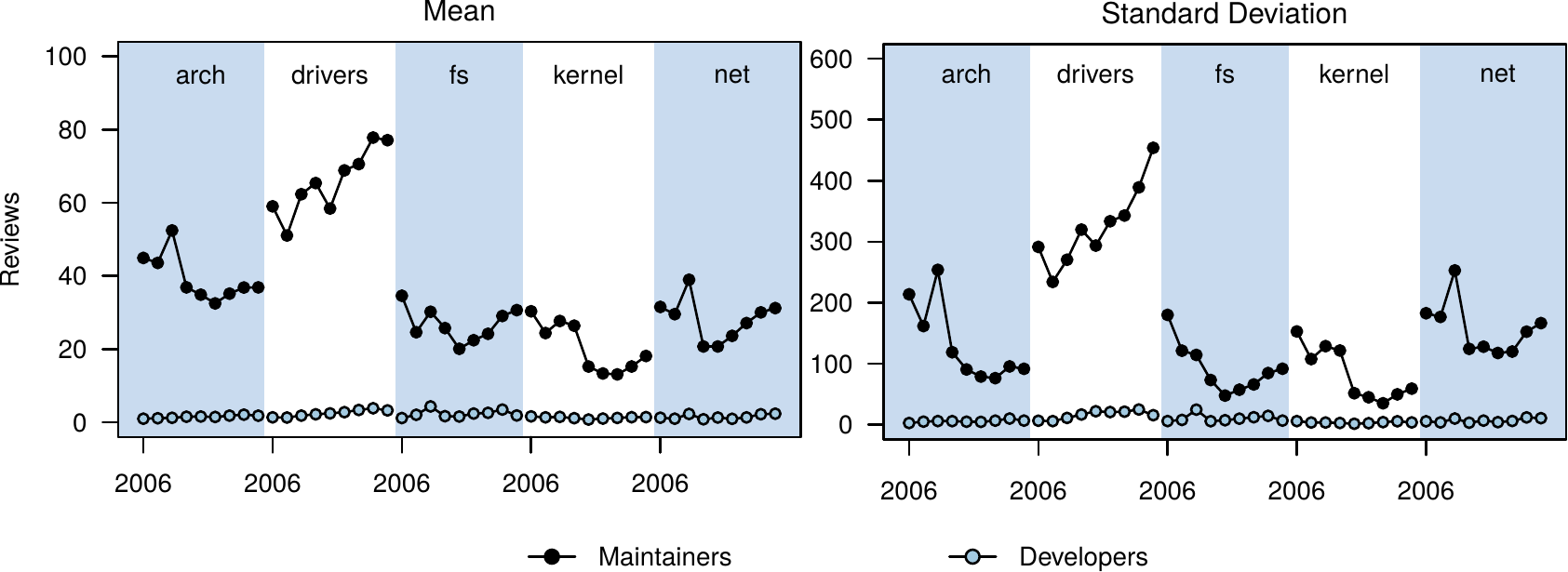}
\caption{Reviews by maintainership (out-strengths)}
\label{fig: maintainers}
\end{figure}

\begin{submittedText}\submittedTextNote{JAIS March 2016}

As can be seen from Fig.~\ref{fig: maintainers}, the average number of reviews is much higher among maintainers compared to the group of normal developers, many of whom have no reviewed commits at all. Also the standard deviations are much higher in the maintainer group\footnote{As early mentioned on section \ref{section: data}, even if maintainers hold formal responsibilities on specific code-sections of a specific subsystem, they can review as well code submitted to other subsystems.}, which indicates that most of the noted extreme outliers refer to maintainers. 
The corresponding average percentage shares are shown in Table~\ref{tab: review ratios mean}. The relative ratios are particularly high in \texttt{arch}, \texttt{drivers}, and \texttt{kernel}, meaning that many maintainers in these subsystems have reviewed other maintainers. It is worth noting that the degree of homophily is rather high in \texttt{drivers}, although the subsystem contains the highest absolute amount of nodes (see \Cref{subfig:number-nodes}
and the lower ratio of maintainers vs. normal developers (see \Cref{subfig:maintainer-shares}). %
In general, the subsystem differences are likely related to different peer review and patch submission practices that are customary to the daily development in the respective subsystems. A more theoretical interpretation can be left open, nevertheless. In terms of hierarchy, it could be, for instance, that particularly reviews in the development of device drivers pass through many nodes along paths that contain different layers of maintenance, sub-maintenance, and development (i.e., in a delegation hierarchy).
\end{submittedText}

\begin{table}[ht]
\centering
\caption{Average maintainer review ratios (\%)}
\label{tab: review ratios mean}
\begin{threeparttable}
\begin{small}
\begin{tabular}{lccccc}
\toprule
$\qquad\quad$ & \texttt{arch} & \texttt{drivers} & \texttt{fs} & \texttt{kernel} & \texttt{net} \\
\midrule
2006 & 29.22 & 32.55 & 29.00 & 21.94 & 27.28 \\ 
2007 & 32.64 & \colorbox{paleaqua}{39.14} & 22.48 & 31.00 & 29.22 \\ 
2008 & \colorbox{paleaqua}{36.88} & \colorbox{paleaqua}{36.15} & \colorbox{paleaqua}{34.81} & 32.40 & 28.72 \\ 
2009 & \colorbox{paleaqua}{34.21} & \colorbox{paleaqua}{33.83} & 30.59 & \colorbox{paleaqua}{36.05} & 26.94 \\ 
2010 & \colorbox{paleaqua}{34.69} & \colorbox{paleaqua}{36.98} & 30.54 & \colorbox{paleaqua}{39.99} & 30.76 \\ 
2011 & \colorbox{paleaqua}{39.01} & \colorbox{paleaqua}{35.21} & 29.70 & 32.14 & 22.77 \\ 
2012 & \colorbox{paleaqua}{38.09} & \colorbox{paleaqua}{34.51} & 27.56 & \colorbox{paleaqua}{41.78} & 28.50 \\ 
2013 & \colorbox{paleaqua}{39.25} & \colorbox{paleaqua}{37.15} & 28.76 & \colorbox{paleaqua}{42.39} & 31.73 \\ 
2014 & \colorbox{paleaqua}{41.01} & \colorbox{paleaqua}{39.29} & 25.18 & \colorbox{paleaqua}{43.44} & 26.12 \\ 
\bottomrule
\end{tabular}
\end{small}
\begin{scriptsize}
\begin{tablenotes}
\item[]{Values larger than or equal the mean share are colored.}
\end{tablenotes}
\end{scriptsize}
\end{threeparttable}
\end{table}
Although it remains debatable what suffices as an acceptable threshold in descriptive statistics, \ref{h: maintainers and maintainers} can be accepted already on the grounds that, on average, in all samples over one fifth of all maintainers have reviewed other maintainers. At this point the cumulative evidence can be also seen as sufficient to reject \ref{h: subsystem similarity} and \ref{h: annual similarity}. There are empirically relevant differences between the sampled peer review networks.

\subsection{Affiliation as a foci of homophily}

The conjecture \ref{h: affiliations} presumes a homophily tendency that developers from large companies tend to review other developers with the same affiliation. There are a couple of difficulties related to the assertion. First, the extraction of affiliations from e-mail addresses comes with the theoretical assumption that developers are affiliated with the organization that own the corresponding e-mail domain. The second issue is more practical: the five subsystems differ considerably in terms of the largest reviewer groups; semiconductor companies do not commit extensively to \texttt{net} within which networking companies often operate; the \texttt{fs} subsystem is of special interest to companies working in the field of storage; and so forth\footnote{The article published by \citet{cass2014s_who_write_linux} in the IEEE Spectrum magazine and institutional reports from the Linux Foundation \citep[see][]{corbet2015_who_writes_linux} provide aggregated figures on who contributes to Linux. Unfortunately, we are unaware documentation reporting on contributions by subsystem.}. Even for the largest reviewer groups, then, annual and cross-subsystem comparisons are difficult to make already because the resulting frequency distributions are small in some subsamples. While keeping this remark in mind, a descriptive evaluation can be carried out by first considering three well-known companies associated with the Linux kernel development: Intel (one of the world's largest semiconductor companies), Red Hat (a well-known Linux vendor), and SUSE (a competitor for Red Hat's products developed historically by Novell, and now owned by the Micro Focus International). The review ratio from the previous section suffices as a descriptive index. To briefly rephrase the meaning: the index gives the (percentage) share of reviews that, for instance, a Red Hat affiliate has done towards other Red Hat affiliates, scaled by the total number of peer reviews carried out by the affiliate.
\begin{figure}[t]
\centering
\includegraphics[width=0.95\textwidth]{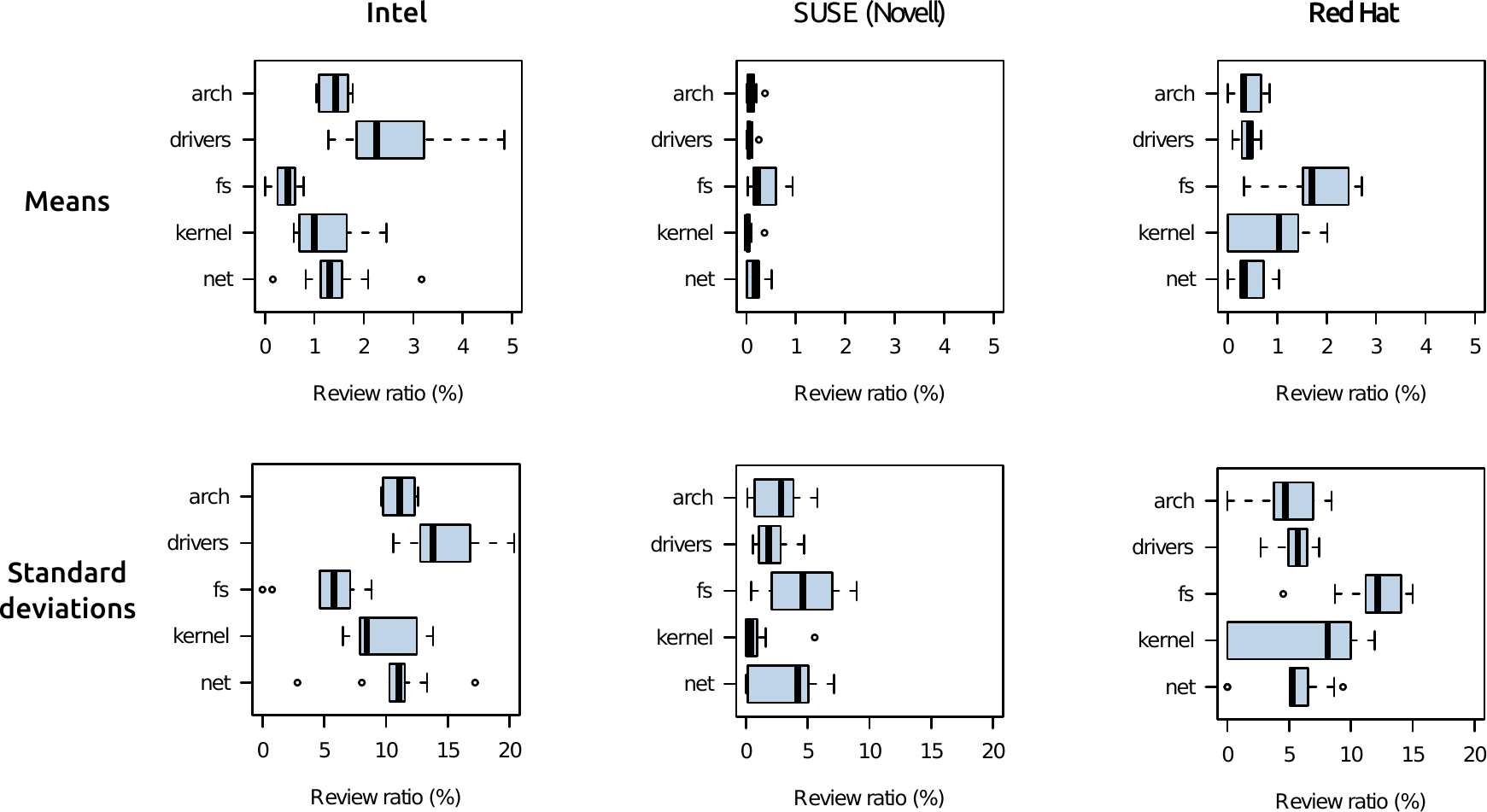}
\caption{Review ratios for three companies (\%)}
\label{fig: review ratios companies}
\end{figure}

\begin{submittedText}\submittedTextNote{JAIS March 2016}
The results are visualized with the six box-and-whisker plots in Fig.~\ref{fig: review ratios companies}. The average review ratios (\%) are shown on the left-hand side, whereas the three plots in the second column show the standard deviations of the percentage ratios. Since the plots are factored according to subsystems, in each plot the conveyed information (namely, central tendency, variance, and outliers) is read annually across the years. For instance, the top-left plot shows that there has been a rather large annual variation in the average review ratios of the Intel affiliates working in \texttt{drivers}. In general, however, the noteworthy detail relates to the scales: for all three affiliation groups, on average less than five percent of all groups' reviews have targeted members with the same affiliations. When looking at the right-hand side plots, it becomes evident that variation is rather large among the individual developers affiliated with the three companies.
\end{submittedText}

\begin{submittedText}\submittedTextNote{JAIS March 2016}

It is evident that the varying top-three affiliation groups exhibit only rather modest average degrees of homophily. These are accompanied by relatively large standard deviations, reflecting heterogeneity among the individual, affiliated, developers. Given that a coarse threshold of 20 \% was interpreted as a sufficient acceptance criterion in the case of maintenance, the analogous homophily conjecture for affiliations, \ref{h: affiliations}, can be tentatively rejected\footnote{With such results (i.e., low average degrees of homophily regarding affiliation), we can reject the conjecture without employing more complex relational methods addressing homophily in evolving social networks such as exponential random graph models (ERGMs) often employed to analyze longitudinal network data.  \citep[see][]{contractor2006testing}.
While ERGMs have been widely used in social network research, they are not yet established  in the IS discipline. Furthermore, they are also computationally intensive and require much computational power for handling large networks.}. It seems that the affiliation attributes do not support a powerful homophily tendency -- at least not when compared to the maintainership attribute.
\end{submittedText}

\section{Discussion}\label{section: discussion}

\begin{submittedText}\submittedTextNote{JAIS March 2016}
This paper approached one research question -- does the code reviews in the Linux Kernel tend to be homophilous?  -- through  network theory, network analysis, \numberOfConjectures\ conjectures, five Linux kernel subsystems, and nine years we provide further insights on 'who reviews who in Linux'. The analytical evaluation model was based on the theory of homophily -- actors are likely to structure their social network according to principles of similarity.  The empirical findings support the theory for a limited extend. 
\end{submittedText}

\subsection{Key findings}

\begin{submittedText}\submittedTextNote{JAIS March 2016}
The empirical results are enumerated in Table~\ref{tab: results}. 
\end{submittedText}

\begin{submittedTable}[JAIS March 2016]
\begin{table}[H]
\centering
\caption{Summary of results}
\label{tab: results}
\begin{threeparttable}
\begin{small}
\begin{tabularx}{11.5cm}{llX}
\toprule
$\textmd{C}_i$ & Support & Description \\
\midrule
\ref{h: maintainers and maintainers} & \colorbox{paleaqua}{Yes} & Maintainers tend to review other maintainers. \\
\cmidrule{3--3}
\ref{h: affiliations} & No & Members affiliated with a large organization only infrequently review  ``colleagues'' affiliated with the same organization.\\
\cmidrule{3--3}
\ref{h: subsystem similarity} & No & The main subsystems differ. \\
\cmidrule{3--3}
\ref{h: annual similarity} & No & Annual variation is large. \\
\bottomrule
\end{tabularx}
\end{small}
\end{threeparttable}
\end{table}
\end{submittedTable}

If prior work along the lines of network science confirmed the presence of core-developers in open-source communities by tracing new code and bugs  \citep{Conaldi13,Crowston06,mockus_et_al2002two_cases}, our analysis confirms the presence of core-reviewers. Such presence is in theoretical accordance with the ideological traits of meritocracy that characterize open-source communities \citep[see][]{o_mahony_and_ferraro2007emergence_governance_oss,stewart_and_gosain2006impact,raymond2001cathedral,parameswaran_and_hinston2007research_on_social_computing}. By assuming that developers become maintainers by merit (i.e., by having a good track record of code-contributions), and that core-developers contribute most of the code \citep{mockus_et_al2000_Apache_case,crowston_and_howison2005health}, it is expectable that core-developers with maintenance responsibilities end up reviewing many contributions from other core-developers.

\begin{submittedText}\submittedTextNote{JAIS March 2016}

The open-source software development communities are also characterized by the co-existence of mechanisms that reinforce both bureaucratic and democratic values \citep{o_mahony_and_ferraro2007emergence_governance_oss,Shaikh2017}.  As different software development roles exist \citep{Conaldi13}, the bureaucratic shared norms that empower maintainers can explain in terms of peer review, the relatively strong homophily, tendency for maintainers -- who review frequently -- to review other maintainers. On this point,  as we found out that "maintainer-role" is a determinant of homophily, the peer review practices of Linux %
convey with the covered theory on social networks and open-source software.
\end{submittedText}

\begin{submittedText}\submittedTextNote{JAIS March 2016}

However, contrary to what was theoretically expected \citep{mcpherson2001_a_review_of_homophily,kossinets_and_watts2009origins_of_homophily,gallivan_and_ahuja2015co}, we found out that the analogous homophily tendency is absent in terms of affiliations. Contrary to postulated by the principle of homophily, we found no evidence of a developers tendency to review other developers with the same organizational affiliation. In such aspect, the Linux peer-review processes remains ``fair'' -- reviewers tend to not review the work of colleagues. This goes in line with the theoretical work of \citet{cooper2005_inclusiveness_and_antirivalry_and_inclusiveness} and \citep{morgan_et_al2013_oss_value_networks} who previously suggested that open-source promotes anti-rivalry and inclusiveness. Contrary to expected from earlier research in social networks \citep{mcpherson2001_a_review_of_homophily,kossinets_and_watts2009origins_of_homophily,gallivan_and_ahuja2015co}, affiliation does not shape the relational patterns of who review who in Linux.  The findings diverge from established research on homophily, but are in line with the few studies that explored so far homophily in the open-source context ~\citep[cf.][]{teixeira_et_al2015lessons,linaaker_et_al2016_coopetition_in_Hadoop}.
If prior related research found that, developers affiliated with competing firms collaborate as openly as developers affiliated with non-rivaling firms do, regardless of their affiliation. Our analysis of peer-review in Linux added then corroborating evidence %
that in terms of relational patterning, it appears that organizational affiliation “does not matter” as much in open-source communities as in other social networks.
\end{submittedText}

\begin{submittedText}\submittedTextNote{JAIS March 2016}

Since we found that ``being a maintainer'' shapes much more strongly the relational pattern of homophily than ``being affiliated with a given organization'', it seems appropriate to emphasize the different software development roles and processes in relation to the peer review practices across the Linux kernel subsystems. It is much more likely that the generative homophily mechanisms are driven by software engineering roles and practices rather than the commercialization trend that affected the Linux kernel development throughout the 2000s. In other words, the firm engagement in the Linux kernel development is unlikely to robustly influence the peer reviewing practice followed in the kernel development.  From a peer review perspective, and besides the powerful social tendency for people to relate to others similar to them,  the ideological values of meritocracy, inclusiveness and non-rivalry seem in good shape besides the Linux commercialization trend.
\end{submittedText}

 \subsection{Contributions}

\begin{submittedText}\submittedTextNote{JAIS March 2016}
Discussed our  key findings, we position our contributions to theory and practice while arguing for the novelty and utility of our research.    
\end{submittedText}

\begin{submittedText}\submittedTextNote{JAIS March 2016}
The case study results largely reflect the prior expectations about peer reviewing in the Linux kernel. The maintainership aspects, in particular, mirror well the documented peer review practices and policies, social hierarchies, and patch submission procedures. Unlike what was presumed, however, in many respects the peer review networks show significant divergence both between the subsystems and annually across time. In analogy between science and open-source software, it is known that scientific peer-review has historically evolved over time \citep[][]{benos_et_al_2007_hir}  and its practices vary from discipline to discipline and journal to journal \citep{cicchetti1991reliability,weller2001editorial_peer_review}.
 In the open-source arena, and in Linux in particular,  our analysis suggests that peer-review practices also evolve over time and vary from subsystem to subsystem. These two aspects undermine the usefulness of the conjectures  \ref{h: subsystem similarity} and \ref{h: annual similarity} for further theoretical hypothesizing.  
\end{submittedText}

After noting that the peer review practices of Linux are highly contextual, it is now worthwhile to summarize the general network tendencies that characterize the peer review networks in the Linux kernel: 

\begin{itemize}
 \item Visible but not uniform homophily across two exogenous node attributes (i.e.,  maintainership and organizational affiliation). 
 \item Maintenance role induces more homophily than organizational affiliation. 
 \item Homophily varies across different subsystems and time. 
\end{itemize}

\begin{submittedText}\submittedTextNote{JAIS March 2016}
These observations have some research implications. As it is important to hypothesize about the generative mechanisms\footnote{{As warned by \cite{eck_et_al2015generative_meaning}, there are different scholarly discourses regarding generativity in IS research. In our research, we refer to mechanisms through which relational structures emerge. We discuss then on the generative mechanisms underlying and producing observable phenomena as in Bhaskar's foundations of critical realism \citep{collier1994critical,bhaskar2013realist}. We do not refer to Zittrain's generativity as outcomes of digital technology \citep[cf.][]{zittrain2006generative,yoo2010research,kallinikos2013}.}}  that drive the evolution of network dynamics \citep[see][for a discussion regarding such theoretical generative mechanisms within socio-material systems]{contractor_et_al2011networks_and_sociomateriality}, %
our research elucidates that the homophily concept offers one high-level theoretical generative mechanism to interpret the evolution of open-source peer review  networks. Givem our results, it remains to be evaluated whether homophily is suitable to crystallize a typical OSS peer review network.
On the other hand, also the network theories that emphasize self-organization, such as the preferential attachment theory and the onion metaphor \citep{Crowston06}, are unlikely to  provide alone a sufficient empirical explanation for the theoretical generation tendencies \citep[see][]{carillo_and_bernard2015many}. The shape of a probability distribution of weighted network degrees is also unlikely to provide further ground for practical optimization models beyond simple hypotheses about correlation with peer review efficiency measures. 
On this point, we concur with the growing recognition among social networks researchers that the emergence of a network can rarely be adequately explained by a single theory \citep{contractor_et_al2011networks_and_sociomateriality,monge_and_contractor2003theories,cederman2005computational_model_in_sociology_process_generative_view,poole_and_contractor2011conceptualizing_teams_as_networked_ecosystem}.   
\end{submittedText}

\ifbool{shortVer}{}{
\begin{submittedText}\submittedTextNote{JAIS March 2016} 
The generative mechanisms also open the door for hypothesizing about future generalizability. There are several directions to try to generalize towards. %
First, 
it is relevant to ask whether the mechanisms generalize to other OSS peer review networks, or whether there is a difference between projects in terms of style of review (review-then-commit vs. commit-then-review) that are widely used by OSS communities \citep{Rigby14,kogut2001open}. 
Second, due to lack  of  physical  interaction many participants  form  impressions  of  their  team-mates  based on offline  vehicles, including: code commits, code reviews, mailing-lists, and bug comments -- further progress is required to systematically access how such vehicles and impressions influence the mechanisms of peer review networks \citep[cf.][]{Bosu14a}. 
Third, it is important to know whether or not, and to which extent, the mechanisms generalize to other OSS collaborative practices (e.g., coding and testing).  
\end{submittedText}

\begin{submittedText}\submittedTextNote{JAIS March 2016}
Forth,%
it is possible to hypothesize about the common characteristics between OSS, software development, and scientific peer review practices. Indeed, homophily along with preferential attachment have been observed to be significant  mechanisms in scientific journal citation networks \citep{Newman01, Peng15}. Then, if the mechanisms are similar, also the problems and possible solutions may be similar. Last, the interdisciplinary network research domain still lacks systematic empirical comparisons between human networks and biological, technical, or other unsocial networks. How the technical network tendencies \citep{Concas07, Ermann11, WangYu13} are related to the mechanisms that generate the social tendencies? At the present state of research it suffices to summarize that only centralization and power-laws allow a straightforward theoretical linkage.
\end{submittedText}

\begin{submittedText}\submittedTextNote{JAIS March 2016}
Fifth, 
regarding the homophily mechanism, 
}

\begin{submittedText}\submittedTextNote{JAIS March 2016}

Our findings add Linux to OpenStack and Hadoop as muddling cases 
where organizational affiliation ``does not matter'' as recently reported \citep[see][]{teixeira_et_al2015lessons,linaaker_et_al2016_coopetition_in_Hadoop}. It is still unknown what is exceptional regarding homophily and organizational affiliation -- the three cases executed so far, or the whole open-source community. Mining large datasets from  software repository/hosting services (e.g., GitHub)
\ifbool{shortVer}{}{
, Bitbucket and Black Duck Open Hub
}
should assess if firms within open-source ecosystems are able to work with possible competitors without rivalry homophilous tendencies. Future research is also required the explain such particular heterophily in open-source cases -- What can explain such low levels of homophily? The norms, beliefs and values of the community?  The virtualization of work practices as developers collaborate with reduced face-to-face interactions? 
The fact that developers identify themselves with the community and not with the organization that they are affiliated with? The fact that developers concentrate their attention in value-creation neglecting value-appropriation? 
\end{submittedText}

\begin{submittedText}\submittedTextNote{JAIS March 2016}
More generally,  and in an attempt to illuminate the directions of IS research, we believe that the discipline can benefit by further overlap with network science. Not only by applying network analysis methods but also by embracing network theory. As information systems become increasingly networked and interconnected \citep{henfridsson_and_bygstad2013generative,ciborra_et_al2000_dynamics_digital_infrastructures} further promising avenues for future research can benefit from both network theory and network analysis. This article attempted to demonstrate such potential by exploring the theoretical principle of homophily and peer-reviewing in the Linux kernel. Future research along these lines could look at peer review in other complex software development projects and explore other mechanisms of network evolution  beyond homophily (e.g, clustering,  preferential attachment, randomness, and small-world phenomena among others). 
\end{submittedText}

\begin{submittedText}\submittedTextNote{JAIS March 2016}
Regarding practical utility, we shed some lights on the Linux peer-review process that should be of practitioners' interesting. After all, Linux remains a reference case -- it is arguably the most studied software project of all time. Moreover, many practitioners are concerned with the commercialization trend in Linux. The so-called ``open-source purists''  consider that the tenets of OSS ideology \citep[see][]{stewart_and_gosain2006impact} are in danger due to the growing corporate involvement in the Linux kernel. 
\citep[e.g.,][]{nyt1999_concerns_of_linux_comercialization,sliwa2004commercialization}. Addressing such concerns, we do not further alarm the purists. When it comes to peer review, our analysis did not found any relational tendency undermining the norms, beliefs and values that characterized Linux since its inception. 
\end{submittedText}

\begin{submittedText}\submittedTextNote{JAIS March 2016}
Finally, by not looking at our outcomes, but rather to the  employed research design. We believe that our method could be applied to investigate many peer-review facets of meritocracy, inclusiveness and rivalry for any given project orchestrated by a software repository - this could be of especial interest for open-source stakeholders wishing to conduct assessments from an ideological perspective. The assessment of community-practices regarding homophily, heterophily, inclusiveness and rivalry could be of governance concern, especially in large scale projects where knowing ``who reviews who" is not as straightforward as in smaller projects. 
\end{submittedText}

\section{Conclusion}

\begin{submittedText}\submittedTextNote{JAIS March 2016}
Based on our finding, we argue that besides the increasing involvement of commercial companies in the development of the Linux kernel, its peer review foundations have been preserved. On one hand, the commercialization trend and the arrival of paid development have not generated a forceful homophily tendency for affiliated developers to review other developers with the same affiliations. A contrary result would have been alarming for the OSS advocates and their ideologies.
\ifbool{shortVer}{}{
On the other hand, there has been a systematic homophily tendency for maintainers to review each other. Since there are also signs that peer reviewing has become slightly less concentrated over the recent years, it seems that the maintenance and review burdens have been balanced by widening the middle-level in the social hierarchy (i.e., increased delegation). 
\end{submittedText}

\begin{submittedText}\submittedTextNote{JAIS March 2016}
The homophily theory is well-suited also for illustrating more fundamental issues in open-source software development peer reviews. A certain degree of homophily is a necessity for ensuring software quality through peer review. If none of the involved peers share relevant abstract attributes, the logical implication would be that expertise and specialization would no longer play a role in peer review. In terms of the Linux kernel, the total lack of shared attributes would make the kernel subsystems meaningless for development; a peer with expertise in embedded processor architectures could robustly and efficiently review peers specialized in file system development, for instance. Clearly, a higher degree of quality and faster review times could be expected in case the file system developers would be reviewing each other. 
\end{submittedText}

\begin{submittedText}\submittedTextNote{JAIS March 2016}
The issue is not comparable to seniority or socialization into the kernel development. open-source maintainership is about code ownership and software maintenance, and it cannot be conjectured that only senior developers would engage themselves in maintenance.
The encouragement and recruitment of newcomers have constituted the practical recommendation to reduce the network centralization tendencies and the core-periphery effects \citep{Bosu14b, Toral10}. However, alternatively, or in addition, code ownership can be increased by promoting maintainership responsibilities. This promotion is observable in the development of Linux device drivers, most of which require highly specialized knowledge, but which may or may not correlate strongly with kernel development seniority as such. Both seniority and maintainership may increase efficiency. Since no forceful homophily tendency is observed at the firm level, maintainership promotion seems safe in terms of the normative OSS peer review ideals -- even though  maintainership may also well offer an easy stepping stone for firm engagement and commercial fortification of open-source code ownership.
\end{submittedText}

The degree of homophily should not reach a maximum, however. }As Linus Torvalds emphasized already in the late 1990s, the point of OSS peer reviews is to find developers with different backgrounds to review each other \citep{lee_and_cole2003oss_model_of_knowledge_creation}. %

\end{submittedText}

\section{Acknowledgments}

\input{./wp-acknowledgments.tex}
\printaddresses

\printbibliography

\renewcommand{\thesection}{A}
\renewcommand{\thetable}{\thesection.\arabic{table}}
\renewcommand{\thefigure}{\thesection.\arabic{figure}}

\end{document}

%% file: wp-meta-data.tex
 \newrobustcmd{\wptitle}{Network Science, Homophily and Who Reviews Who in the Linux Kernel?}
 \newrobustcmd{\wpsubtitle}{Network Science, Homophily and Peer Review in Linux}

\newrobustcmd{\wpasin}{As presented at 2020 European Conference on Information Systems (ECIS 2020), held Online,  June 15-17, 2020.  The official conference proceedings are available at the  AIS eLibrary (\url{ https://aisel.aisnet.org/ecis2020_rp/}).}

\newrobustcmd{\wprunningauthors}{Jose Apolinário Teixeira et al.}

\newrobustcmd{\wpsubject}{28th European Conference on Information Systems (ECIS 2020)}

\newrobustcmd{\wpkeywordslist}{Peer Review, Open-source, Network Science, Homophily, Linux}

\newrobustcmd{\wppaurl}{\url{http://users.abo.fi/jteixeir/pub/linuxsna/ECIS2020-arxiv.pdf}}

%% file: wp-template.head.tex
\usepackage[usenames,dvipsnames]{xcolor}

\usepackage[utf8]{inputenc} %
\usepackage[english]{babel} %

\usepackage{csquotes}

\usepackage{anyfontsize}

\usepackage[T1]{fontenc} %

 \usepackage{stix}
 \usepackage{libertine,libertinust1math} %

\newcommand\titlepagedecoration{%
\begin{tikzpicture}[remember picture,overlay,shorten >= -10pt]

\coordinate (aux1) at ([yshift=-15pt]current page.north east);
\coordinate (aux2) at ([yshift=-410pt]current page.north east);
\coordinate (aux3) at ([xshift=-4.5cm]current page.north east);
\coordinate (aux4) at ([yshift=-150pt]current page.north east);

\begin{scope}[titlepagecolor!40,line width=12pt,rounded corners=12pt]
\draw
  (aux1) -- coordinate (a)
  ++(225:5) --
  ++(-45:5.1) coordinate (b);
\draw[shorten <= -10pt]
  (aux3) --
  (a) --
  (aux1);
\draw[opacity=0.6,,shorten <= -10pt]
  (b) --
  ++(225:2.2) --
  ++(-45:2.2);
\end{scope}
\draw[titlepagecolor,line width=8pt,rounded corners=8pt,shorten <= -10pt]
  (aux4) --
  ++(225:0.8) --
  ++(-45:0.8);
\begin{scope}[titlepagecolor!70,line width=6pt,rounded corners=8pt]
\draw[shorten <= -10pt]
  (aux2) --
  ++(225:3) coordinate[pos=0.45] (c) --
  ++(-45:3.1);
\draw
  (aux2) --
  (c) --
  ++(135:2.5) --
  ++(45:2.5) --
  ++(-45:2.5) coordinate[pos=0.3] (d);   
\draw 
  (d) -- +(45:1);
\end{scope}
\end{tikzpicture}%
}

\usepackage{textcomp}

\usepackage{pifont}

\usepackage{float}

\usepackage[caption=false]{subfig}

\usepackage{fancyhdr}

\usepackage{epigraph}

\usepackage{adjustbox}

\usepackage{blindtext} %

\usepackage[activate={true,nocompatibility},final,tracking=true,kerning=true,spacing=true,factor=1100,stretch=10,shrink=10]{microtype}

\usepackage{array}
\usepackage{booktabs}

\usepackage{tabularx} %

\usepackage[flushleft]{threeparttable}

\usepackage{enumitem} %
\setlist[itemize]{noitemsep} %

\usepackage[]{enumitem}

\binoppenalty=\maxdimen
\relpenalty=\maxdimen

\usepackage{epigraph}

\usepackage{lettrine} %

\usepackage{titlesec} %
\renewcommand\thesection{\Roman{section}} %
\titleformat{\section}[block]{\large\centering\bfseries}{\thesection.}{1em}{\vspace{0.3em}} %
\titleformat{\subsection}[block]{\normalfont\large\bfseries}{}{0.0em}{\vspace{0.15em}} %

\makeatletter
\appto\@floatboxreset{%
  \ifx\@captype\andy@table
    \normalfont
  \fi
}
\def\andy@table{table}
\makeatother

\definecolor{titlepagecolor}{cmyk}{1,.60,0,.40}

\DeclareFixedFont{\titlefont}{T1}{ppl}{b}{n}{0.28in}
\DeclareFixedFont{\subtitlefont}{T1}{ppl}{b}{it}{0.22in}
\DeclareFixedFont{\workingpaperfont}{OT1}{cmss}{m}{n}{0.21 in}

\DeclareFixedFont{\footnotefont}{OT1}{cmss}{m}{n}{0.05 in}
\DeclareFixedFont{\figCaptionfont}{OT1}{cmss}{m}{n}{0.13 in}

\usepackage{fontawesome5}

\newcommand{\CRornamentleft}{%
{\color{joseViolet2016}\faIcon{file-alt}}
}
\newcommand{\CRornamentright}{%
{\color{joseViolet2016}\faIcon{file-alt}}
}

\newcommand{\CRornamentheader}[1]{%
    \vspace{0.3cm} 
    \begin{center}
    \CRornamentleft
    \hspace{4pt} {\large\emph{#1}}\hspace{4pt}  %
    \CRornamentright
    \end{center}%
    \vspace{0.3cm} 
}

\newcommand{\Fornamentheader}[1]{%
\vspace{0.3cm} 
    \begin{center}
    {\color{joseViolet2016} \faKeyboard[regular]}
    \hspace{4pt} {\large\emph{#1}}\hspace{4pt}   %
    { \color{joseViolet2016}  \faKeyboard[regular]}
    \end{center}%
    {\vspace{0.3cm}}
}

\newcommand{\HISTornamentheader}[1]{%
    \begin{center}
    {\color{joseViolet2016}  \faFeather}
    \hspace{4pt} {\large\emph{#1}}\hspace{4pt}   %
    {\color{joseViolet2016}  \faFeather}
    \end{center}%
    
}

\newcommand{\FUNDornamentheader}[1]{%
\vspace{0.3cm} 
    \begin{center}
    { \color{joseViolet2016} \faInfoCircle} 
    \hspace{4pt} {\large\emph{#1}}\hspace{4pt}   %
    { \color{joseViolet2016} \faInfoCircle}
    \end{center}%
    \vspace{0.3cm} 
}

%% file: wp-article-header.tex
\input{./config.tex}
\input{./latextoolsforwritingacademicpapers/reviewAndCommentTools.tex}

\input{./latextoolsforwritingacademicpapers/submissionAndRevisionTools.tex}

\input{bibliographies.config.tex}

\makeatletter
\let\c@author\relax
\makeatother

\usepackage[backend=biber,style=authoryear,natbib=true]{biblatex}
\ifbool{final}
{
    \hypersetup{
    pdfinfo={
        Author = {Jose Apolinário Teixeira et al.},
        Title ={Network Science, Homophily and Who Reviews Who in the Linux Kernel?}, 
        Subject = {28th European Conference on Information Systems (ECIS 2020)},
        Keywords = {Peer Review, Open-source, Network Science, Homophily, Linux}
  }
}
 
}
{
  \hypersetup{
    colorlinks = true,
    citecolor=black,
    }
}

\definecolor{paleaqua}{rgb}{0.74, 0.83, 0.9}

\newcommand{\mta}{\textup{\textbf{\textrm{A}}}}

\newcommand{\numberOfConjectures}{four}

\ifbool{final}
{

 \renewcommand{\markabove}[2]{}
 \renewcommand{\markbelow}[2]{}
 \renewcommand{\boxreview}[2]{}
}
{}

\iftoggle{boxSubmittedContent}{%
  \newtcolorbox{submittedText}[1][]{
  oversize,
  breakable,
  enhanced,
  before upper={\parindent15pt},
  colframe=Melon,
  colback=Melon!10
  #1}
  }
  { %
  \newenvironment{submittedText}{}{}
  }
 
\newtcolorbox[]{submittedNote}[1][]
  {title= \footnotesize {Under review},
   width=3cm,
   left=0pt,
   right=0pt,
   fonttitle=\bfseries\color{Brown},
   colframe=Melon,
   colback=Melon!10,
   #1
}

\newcommand\submittedTextNote[1]{

  \iftoggle{boxSubmittedContent}{%

    \marginnote[]{%
    \makebox[0pt][l]{\begin{submittedNote}
    \footnotesize \centering Submitted to #1 
    \end{submittedNote}}}%

  }
  
  {%
  }

}

%% file: config.tex
\usepackage{etoolbox}

\newtoggle{boxSubmittedContent}

\togglefalse{boxSubmittedContent}

\newbool{final}

\booltrue{final}

\newtoggle{draftVersion}

\togglefalse{draftVersion}

\newbool{shortVer}

\booltrue{shortVer}

%% file: latextoolsforwritingacademicpapers/submissionAndRevisionTools.tex
\usepackage[dvipsnames]{xcolor}
\usepackage[many]{tcolorbox}
\usepackage{marginnote}
\usepackage{framed}
\usepackage{float}

\newtcolorbox[]{subAndRevbox}[3][]
{
  colframe = #2,
  colback  = #2!10,
  coltitle = #2!20!black,  
  fonttitle=\bfseries\color{Brown},
  title    = \footnotesize  #3,
  #1,
}

\newenvironment{submittedTable}[1][Submitted somewhere]
    {  %
      \iftoggle{boxSubmittedContent}{%
      \centering 
      \begin{minipage}{1.0\textwidth}
      \begin{subAndRevbox}{Melon}{ \centering Table as submitted to #1}
      }
      {%
      }

  }
  {  %
  
       \iftoggle{boxSubmittedContent}{%

      \end{subAndRevbox} %
      \end{minipage}
       
      }
      {%
      }
}

%% file: wp-hard-cover.tex
\pagenumbering{roman} %
\cfoot{\thepage} %

\renewcommand{\thefootnote}{\ding{80}} 

\makeatletter                       
\def\printauthor{%
    {\small \@author}}              
\makeatother
\author{%
    José Apolinário Teixeira \\
    Åbo Akademi University \\ Finland \\
    { \small   \faMailBulk \ \textbf{jteixeira@abo.fi}}
    \vspace{5pt} \\
     Ville Leppänen \\ 
    University of Turku \\ Finland \\ 
    \vspace{5pt} 
    Sami Hyrynsalmi \\ 
    LUT University \\  Finland \\
    }

\begin{figure}[t]
 \centering
 \includegraphics[keepaspectratio=true,width=0.85\textwidth]{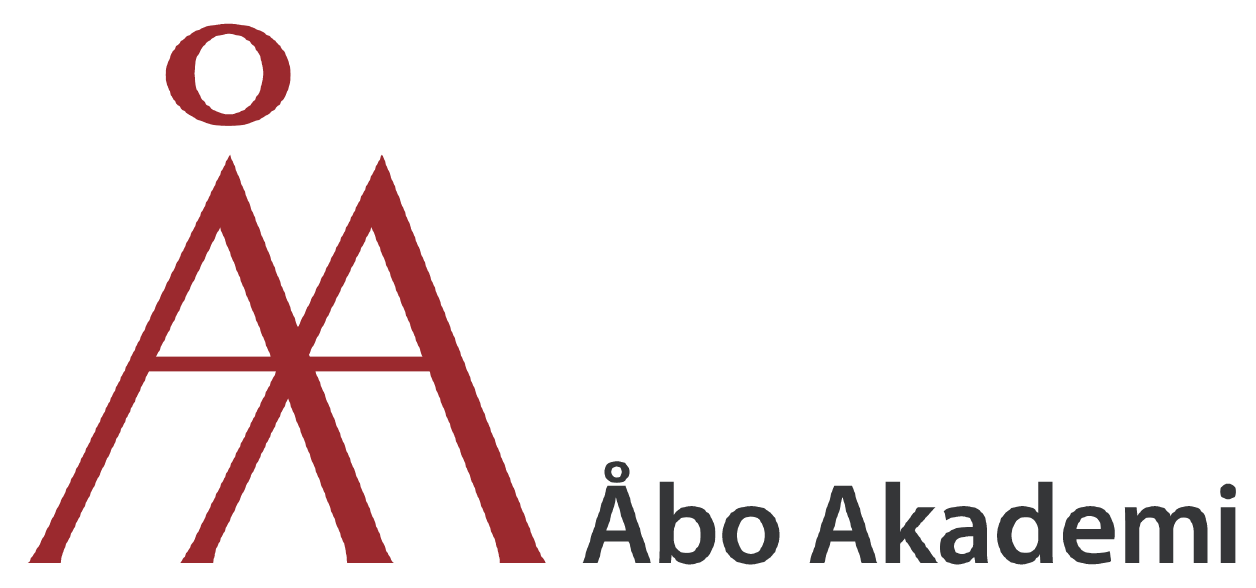}
\end{figure}

\vspace{0.5cm}

\vspace{0.5cm}

\begin{titlepage}

\vspace*{1cm}

\noindent   \titlefont \wptitle %

\noindent  \workingpaperfont Working paper\footnote{\wpasin
 }
 
 \vspace{0.2cm}

 \noindent  \normalfont { \small \color{titlepagecolor} \faUnlock* } Open-access at \wppaurl
 
\vspace{1.5cm}

\vspace*{1cm}
\noindent
\hfill
\begin{minipage}{0.80\linewidth}
    \begin{flushright}
        \printauthor
    \end{flushright}
\end{minipage}
\begin{minipage}{0.02\linewidth}
    \rule{1pt}{120pt}
\end{minipage}
\titlepagedecoration

\end{titlepage}

\vfill
\newpage

\normalfont

\normalfont
\CRornamentheader{Copyright notice}

 The copyright is held by the authors. 
 The same article is available at the \href{https://aisel.aisnet.org/ecis2020_rp/}{AIS Electronic Library (AISeL)} with permission from the authors ({\small see \url{https://aisel.aisnet.org/ecis2020_rp/}}).  The Association for Information Systems (AIS) can publish and reproduce this article as part of the  Proceedings of the European Conference on Information Systems (ECIS 2020).

\CRornamentheader{Archiving information}

The article was self-archived by the first author at its own personal website \wppaurl \ during June 2021 after the work was presented at the 28th European Conference on Information Systems (ECIS 2020).

\vfill
\newpage

\normalfont

\FUNDornamentheader{Funding and Acknowledgements}

\input{./wp-acknowledgments.tex}

\vfill
\newpage

\normalfont

\Fornamentheader{Formatting and typesetting}

This working paper was formatted in Latex\footnote{Version details: pdfTeX, Version 3.14159265-2.6-1.40.15 (TeX Live 2015/dev/Debian)} and its available in PDF v.1.5 with rich metadata. It is based on the 'working papers' template from the first author. Special thanks to Vytas Statulevicius from VTeX and Jacky Lee from the Chinese University of Hong Kong for 'prior art' on the SpringerOpen BMC template. 

\clearpage
{}   \HISTornamentheader{Abstract in the English language} {  }

 \vspace{0.3cm} 

\input{./abstract.tex}

 \vspace{0.3cm}

 \clearpage
 \HISTornamentheader{Abstract in the Portuguese language}
 \vspace{0.3cm} 

 \renewcommand{\abstractname}{Resumo}

 \begin{abstract}
\vspace{0.5cm}

  A revisão por pares é uma prática comum de controlo de qualidade tanto na ciência quanto no desenvolvimento de software.
Nesta pesquisa, investigamos a revisão por pares no desenvolvimento do sistema operativo Linux usando teoria e métodos para 
 a análise de redes sociais. Estruturamos um modelo analítico que integra o princípio sociológico da homofilia (ou seja, a tendência relacional de cada individual para estabelecer relações com outros semelhantes a si) no contexto de revisão por pares no desenvolvimento de software de código aberto em particular. 
 Encontramos uma tendência relativamente forte de homofilia para os mantenedores de revisar outros mantenedores, mas uma tendência comparável está surpreendentemente ausente em relação à afiliação organizacional dos diferentes programadores. Tais resultados reflectem as normas, crenças, valores, processos, políticas e hierarquias sociais documentadas que caracterizam o desenvolvimento do kernel Linux. Os nossos resultados realçam o valor da teoria da análise de redes sociais para explicar a evolução da revisão por pares no desenvolvimento de software.  Em relação à preocupação dos profissionais sobre a tendência de comercialização do Linux, nenhuma tendência de programadores para revisar programadores afiliados com a mesma organização  (colegas profissionais) for encontrada.

 \end{abstract}

  \renewcommand{\abstractname}{Abstract}

 \clearpage

\clearpage %
\pagenumbering{arabic} %
\cfoot{\thepage\ of \lastpageref{pagesLTS.arabic}}

 \renewcommand*{\thefootnote}{\arabic{footnote}}

 \setcounter{footnote}{0}

%% file: wp-acknowledgments.tex
The first author's efforts were partially financed by \textit{Liikesivistysrahasto} - the Finnish Foundation for Economic Education, the \textit{Academy of Finland} via the DiWIL project (see \href{http://abo.fi/diwil}{http://abo.fi/diwil}) project.
 A research companion website at \href{http://users.abo.fi/jteixeir/ECIS2020cw}{http://users.abo.fi/jteixeir/ECIS2020cw}  supports the paper with additional methodological details, additional data visualizations (plots, tables, and networks), as well as high-resolution versions of the figures embedded in the paper. Also, the same website pinpoints some limitations of our approach and outlines as well a promising avenue for future research to further investigate peer review in the context of software development.

%% file: abstract.tex
\begin{abstract}

Peer review is a common quality control practice in both science and software development.
In this research, we investigate peer review in the development of Linux by drawing on network theory and network analysis. 
We frame an analytical model which integrates the sociological principle of homophily (i.e., the relational tendency of individuals to 
establish relationships with similar others) with prior research on peer-review in general and open-source software in particular. 
We found a relatively strong homophily tendency for maintainers to review other maintainers, but a comparable tendency is surprisingly absent
regarding developers' organizational affiliation. Such results mirror the documented norms, beliefs, values, processes, policies, and social hierarchies that characterize the Linux kernel development. 
Our results underline the power of generative mechanisms from network theory to explain the evolution of  peer review networks. 
Regarding practitioners' concern over the Linux commercialization trend, no relational bias in peer review was found albeit the 
increasing involvement of firms. 

\end{abstract}

%% file: wp-article-first-page.tex
\begin{frontmatter}

\begin{fmbox}
\dochead{Research working paper}

\title{\wptitle\thanks{\footnotesize \wpasin}}

\author[
   addressref={aff1},                   %
   corref={aff1},                       %
   email={jose.teixeira@abo.fi}   %
]{\inits{JAT}\fnm{Jose} \snm{Apolinário Teixeira}}
 \author[
    addressref={aff2},
 ]{\inits{VL}\fnm{Ville} \snm{Ville Leppänen}}
\author[
    addressref={aff3},
]{\inits{SH}\fnm{Sami} \snm{Hyrynsalmi}}

\address[id=aff1]{%
  \orgname{Åbo Akademi University}, %
  \street{Domkyrkotorget 3},                     %
  \postcode{FI-20500}                                %
  \city{Åbo},                              %
  \cny{Finland}                                    %
}
\address[id=aff2]{%
  \orgname{University of Turku},
  \street{Turun yliopisto}
  \postcode{FI-20014}, 
  \city{Turku},
  \cny{Finland}
}
\address[id=aff3]{%
    \street{Yliopistonkatu 34}
    \postcode{FI-53851}, 
    \city{Lappeenranta},
  \orgname{LUT University},
  \cny{Finland}
}

\begin{artnotes}
\end{artnotes}

\end{fmbox}%

\begin{abstractbox}

\input{./abstract.tex}
\begin{keyword}
\kwd{Peer Review}
\kwd{Open-source}
\kwd{Network Science}
\kwd{Homophily}
\kwd{Linux}
\end{keyword}

\end{abstractbox}

\end{frontmatter}